\documentclass[preprint,aps,showpacs]{revtex4-1}

\usepackage{amsmath}
\usepackage{graphicx}
\usepackage{bm}
\usepackage[breaklinks=true]{hyperref}
\usepackage[titletoc]{appendix}
\usepackage{color}
\usepackage{subfig}
\usepackage{ulem}

\input epsf.sty

\begin{document}

\title{Crystallization in Glassy Suspensions of Hard Ellipsoids} 

\author{Sven Dorosz}
\affiliation{Theory of Soft Condensed Matter, Universit\'e du Luxembourg, L-1511 Luxembourg, Luxembourg}
\email{sven.dorosz@uni.lu}
\author{Tanja Schilling}
\affiliation{Theory of Soft Condensed Matter, Universit\'e du Luxembourg, L-1511 Luxembourg, Luxembourg}

\begin{abstract}
We have carried out computer simulations of overcompressed suspensions of 
hard monodisperse ellipsoids and observed their crystallization dynamics. 
The system was compressed very rapidly in order to reach the regime of 
slow, glass-like dynamics. We find that, although particle dynamics become 
sub-diffusive and the intermediate scattering function clearly develops a 
shoulder, crystallization proceeds via the usual scenario: nucleation and 
growth for small supersaturations, spinodal decomposition for large 
supersaturations.

In particular, we compared the mobility of the particles in the regions where 
crystallization set in with the mobility in the rest of the system. We did not 
find any signature in the dynamics of the melt that pointed towards the 
imminent crystallization events.  
\end{abstract}

\maketitle
\section{Introduction}

A liquid that is cooled or overcompressed beyond its freezing point 
either crystallizes or forms a glass. If the degree of undercooling 
(or overcompression, respectively) is small, one usually observes crystallization via nucleation 
and growth. If the system is quenched beyond its 
spinodal, it crystallizes immediately. These crystallization mechanisms 
compete with another possible process, the glass transition, which occurs 
in many materials if they are quenched sufficiently fast \cite{Chaikin2000,Binder2011}. 
And between these extremes, mixed routes to crystallization appear,
which can, in general, not be described in terms of a simple free 
energy landscape picture .

A supersaturated melt close to the glass transition resembles 
a liquid in structure, but it differs from 
a liquid in its dynamical behaviour. In particular, on approach to 
the glass transition a melt displays
``dynamic heterogeneity'', i.e.~spatial fluctuations in its 
local dynamical behaviour \cite{Berthier2011a, Berthier2011b}. 

In this article we would like to address the question whether dynamic 
heterogeneity and crystallization are linked. One could e.g.~assume that
regions of enhanced mobility are more likely to crystallize than regions of 
reduced mobility (because the system samples its local phase space more 
rapidly in regions of enhanced mobility) and hence attempt to predict 
positions and times of future 
crystallization events from the spatio temporal structure of the melt.

As a model system we chose mono-disperse hard ellipsoids of revolution 
(spheroids). This system has been shown to exhibit a glass 
transition\cite{Pfleiderer2008, LetzLatz99}, caused by the orientational degree of freedom of the ellipsoids, which acts as a 
source of disorder that is 
sufficiently strong to supress the crystallization process.
On the other hand, the interaction 
potential is very simple (only hard body exclusion) and the system is
mono-disperse, therefore a large part of the equilibrium phase diagram 
has also been mapped out\cite{Frenkel1985, Odriozola2012, Radu2009, Pfleiderer2007, Bezrukov2006,Talbot1990, Donev2004b}. 
Many of the model systems used in other, similar studies either 
show a glass transition or have a simple 
equilibrium phase diagram, but they do not fulfill both requirements at 
the same time. 

The interplay between crystallization and the approach to the 
glass transition in colloids has 
been addressed by several groups recently, but no definite conclusion on 
the mechanism has been drawn \cite{Zaccarelli2009, Pusey2009, Valeriani2011, Sanz2011, Konishi2007, Kawasaki2010}.

\section{Simulation technique and data analysis}

The equilibrium properties of a suspension of hard ellipsoids of revolution 
(spheroids) are determined by the volume fraction $\eta$ and the ellipsoid 
aspect ratio. We will focus on a moderate aspect ratio for which the system 
undergoes a first order phase transition from the fluid phase to a rotator 
crystal phase, i.e.~to a phase with crystalline order in particle positions 
but without orientational order of the axes. In the limit of vanishing 
asphericity the coexistence densities converge to the liquid-solid 
coexistence densities of hard spheres. A slight elongation of the particles 
does not affect the symmetry of the crystal and, as we will show later, the nucleation process also remains unchanged. However, the additional degree of 
freedom acts as a sufficiently strong source of disorder to introduce a glass 
transition \cite{Pfleiderer2008, LetzLatz99}.  

We carried out Monte Carlo (MC) simulations of suspensions of monodisperse hard ellipsoids at constant number of particles $N$ and constant external pressure $P$ (the temperature $T$ is constant, too, but as the system is athermal $T$ only enters the discussion as a trivial factor). Particles were propagated by local translation and rotation moves. The maximum 
displacement per MC step was set to
$0.03$ particle diameters, the maximum rotation of the particle axis to $1.8^\circ$. For such small steps, the ``dynamics'' of the system is similar to 
Brownian dynamics and produces the same behaviour on long time-scales\cite{Berthier2007, Patti2012, Sanz2010}. The system consisted of $N=10386$ prolate hard ellipsoids with an aspect ratio of $a/b=1.25$, where $a$ is the length of the axis of symmetry and $b$ is the length of any axis in the perpendicular plane. To simplify notation, we introduce the dimensionless pressure $P^* = P \frac{8ab^2}{k_BT}$ (where $k_B$ is Boltzmann's constant).  The coexistence pressure is $P^*=14.34$, the coexistence volume fraction of the fluid is $\eta^{\rm coex}_f=0.515$ and the coexistence volume fraction of the crystal is $\eta^{\rm coex}_c=0.544$\cite{Frenkel1985}.

\begin{figure}[h]
\centering
  \includegraphics[height=5cm]{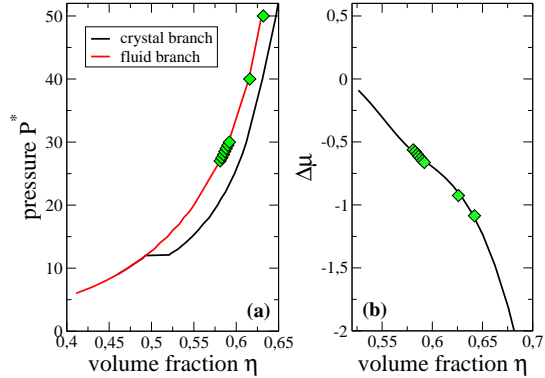}
  \caption{\label{eos} (a) Equation of state for prolate ellipsoids, aspect ratio $a/b=1.25$. (b) Chemical potential difference $\Delta\mu$  between the overcompressed melt and the stable crystal. The diamonds indicate the overcompressions for which we studied the crystallization process.}
\end{figure}

We studied suspensions at constant external pressures $P^* = 27 \ldots 30$, $P^*=40$ and $P^*=50$.  The corresponding chemical potential differences between the stable crystal phase and the meta stable fluid phase $\Delta \mu (P^*)$ have been obtained by integrating along the metastable fluid branch and the stable crystal branch of the equation of state from $P^*_{\rm coex}$ to $P^*$. The values range from $|\Delta\mu|= 0.57\,\frac{k_BT}{\rm particle}$ to $|\Delta\mu|=1.08\,\frac{k_BT}{\rm particle}$. In FIG.~\ref{eos} we illustrate the state points studied here (diamonds) within data which we obtained in MC simulations for (a) the equation of state for our system and (b) the chemical potential difference between the overcompressed melt and the stable crystal.

During the simulation we monitored the volume fraction $\eta$
and the local $q_6q_6$-bond orientational order parameter 
\cite{Steinhardt1983,tenWolde1995}.
For an ellipsoid $i$ with $n(i)$ neighbours, the relative local bond 
orientation is characterized by
\[
\bar{q}_{lm}(i) := \frac{1}{n(i)}\sum_{j=1}^{n(i)} Y_{lm}\left(\vec{r}_{ij}\right)\quad ,
\]
where $ Y_{lm}\left(\vec{r}_{ij}\right)$ are the spherical harmonics and 
$\vec{r}_{ij}$ is the position vector between 
ellipsoid $i$ and its neighbor $j$. In order to identify local fcc-, hcp- or 
rcp-structures l is set to 6. 
A vector $\vec{q}_{6}(i)$ is assigned to each ellipsoid, the elements 
$m=-l \dots l$  of which are defined as 
\begin{equation}
q_{lm}(i) := \frac{\bar{q}_{lm}(i)}{\left(\sum_{m=-l}^l|\bar{q}_{lm}(i)|\right)^{1/2}} \quad .\label{BOP}
\end{equation}

We call an ellipsoid $i$ ``crystalline'', if it has more than eight neighbours $n_b$ with $r_{ij} < 1.5\;b$ and $\vec{q_6}(i)\cdot \vec{q_6}^*(j) > 0.7$. We computed the distributions of $\vec{q_6}(i)\cdot \vec{q_6}^*(j)$ in the bulk liquid, the bulk crystal and for crystallites embedded in a liquid to verify that this criterion, albeit isotropic, works in a solution of anisotropic particles. As the aspect ratio is moderate and as we quench into the rotator phase, the $q_6q_6$-method suffices to detect the crystallites.

For each value of the pressure we examined 25 simulation trajectories that started out from independent starting configurations. 

The ensemble of starting configurations has been prepared in the following way:
All preparation runs were initialized with the same liquid configuration at $\eta=0.567$. This configuration was subjected to an instantaneous increase in pressure and then observed. After $10^6$ MC sweeps, all preparation trajectories had reached a plateau in density and preparation was finished. Although they had started out from the same low density configuration, they had been decorrelated during the $10^6$ MC sweeps of the compression runs. The ensemble of compressed configurations was then used as the starting ensemble for the crystallization study.
The fact that the compressed configurations were decorrelated can be seen e.g.~from the scatter plot of cystal nucleation times FIG.~\ref{fig:scatter} (details of this graph will be discussed further below). We also inspected the system configurations visually. The crystallites always appeared in different places and they had different morphologies.

\section{Results}

\subsection{Dynamics of the supersaturated melt}

\begin{figure}[h]
\centering
  \includegraphics[height=5cm]{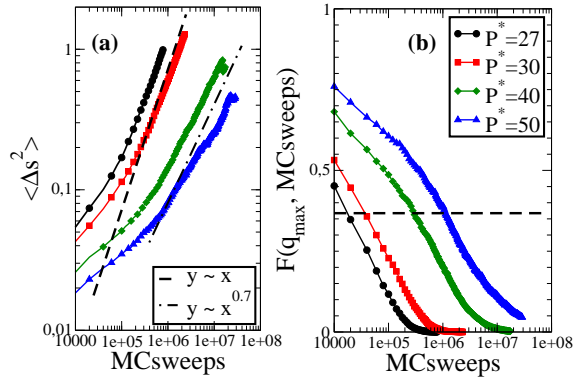}
  \caption{\label{fig1}(a) Mean-squared displacement $\langle\Delta s^2\rangle$ as a function of simulation time for different values of pressure $P^*$. In addition a linear growth law and a sub diffusive power law are plotted for comparison. (b) Dynamic structure factor as a function of simulation time for different values of $P^*$. $q_{{\rm max}}$ corresponds to the first maximum of the static structure factor $S(q)$. The averages are taken over 25 independent simulation runs each. The dashed line marks $1/e$ to indicate the relaxation times.}
\end{figure}

FIG.~\ref{fig1} (left panel) shows the 
average mean-squared displacement of the particles in the melt
\[
\langle \Delta s^2(t) \rangle = \langle \left(\vec{r}_i(t) - 
\vec{r}_i(0)\right)^2\rangle
\]
where $\vec{r}_i(t)$ is the position of particle $i$ at time $t$. The
average $\langle \ldots \rangle$ is taken over all trajectories for a given pressure, and in each trajectory only over particles that are ``liquid-like''
i.e.~not crystalline according to their $q_6q_6$-bond order. The results are robust against our choice of crystallinity parameters, i.e.~variations of the threshold values to identify crystalline particles do not affect the average mean squared displacement since the overall fraction of crystalline particles in the system is small. (We have labelled the distance travelled as $s$ rather than $r$ because we will later introduce 
another quantity that we call $\Delta r$.)
We also computed the self part of the intermediate scattering function of the liquid-like particles 

\[
F_s(q,t)= \left\langle\frac{1}{N}\sum^N_{i}\exp[\textrm{i}\vec{q}(\vec{r}_i(t)-\vec{r}_i(0))]\right\rangle.
\]

From the data shown in FIG.~\ref{fig1} we infer that the simulation runs at the lower values of pressure, $P^*\leq30$, are ``equilibrated'' in the meta-stable liquid basin: The mean squared displacement is linear as a function of simulation time, and the local relaxation times are short in comparison to the induction time for crystal growth (see FIG.~\ref{time_evolution}, discussion follows below). At $P^*\geq 40$ the system is far from equilibrium. The mean-squared displacement is sub-diffusive and the dynamic structure factor decays according to the stretched exponential behaviour that is typical for glass forming systems. (For information on even higher overcompression, we refer the reader to ref.~\cite{Pfleiderer2008}.)

\subsection{Crystallization process}

\begin{figure}[h]
\centering
  \includegraphics[height=5cm]{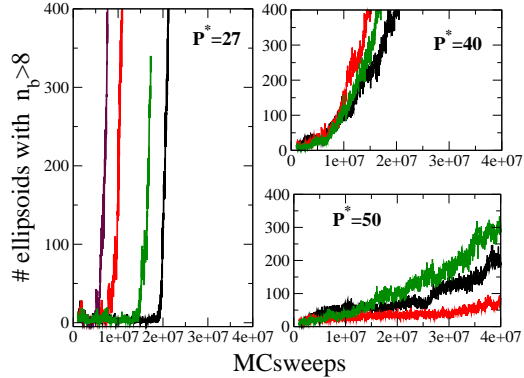}
  \caption{\label{time_evolution} Time evolution of the total number of crystalline particles for different external pressures $P^*$. The lines correspond to individual sample trajectories.}
\end{figure}

The time evolution of the total number of crystalline particles is presented in FIG.~\ref{time_evolution} for several typical simulation runs. In the case of $P^*=27$, there is a long induction time after which the total number of crystalline particles grows rapidly. Particles diffuse over several times their diameter before crystallization sets in. Here we are clearly dealing with nucleation and growth. 
In the case of $P^*=40$ particles diffuse only a fraction of their diameter before crystallization sets in. And at $P^*=50$, the free volume per particle is too small to allow for successful rearrangements on the time scales of our simulations. Here, in the majority of the simulation runs, we do not observe the formation of a crystal.

The main question that we address in this study is whether 
regions of enhanced mobility in the melt are correlated with future 
crystallization sites. However, before we come to this point (in 
subsection \ref{DynCryst}), we would like present some more information on the 
crystallization processes that we observe.  

\begin{figure}[h]
\centering
  \includegraphics[height=4.5cm]{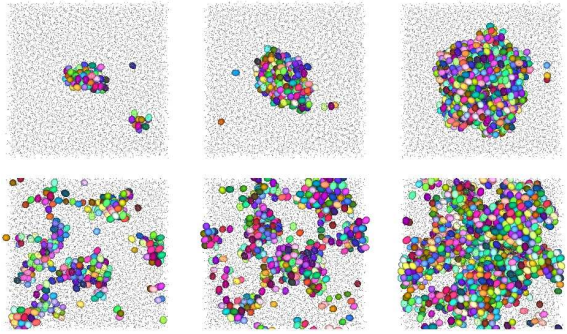}
  \caption{\label{snapshot} Two sequences of snapshots of the crystallization process.  Upper row: $P^*= 27$ at times $1.67 \cdot 10^7$, $1.79\cdot 10^7$ and $1.92\cdot 10^7$ MC sweeps. Lower row: $P^*= 40$ at times $1.08\cdot 10^7$, $1.24\cdot 10^7$ and $1.90\cdot 10^7$ MC sweeps. The ratio of crystalline particles in the configurations is: $0.7\%$, $2.2\%$, $12\%$ (upper row) and $3.2\%$, $5.7\%$, $16.5\%$ (lower row). Liquid-like ellipsoids are shown as dots only.}
\end{figure}

FIG.~\ref{snapshot} shows typical sequences of snapshots.
For $P^*=27$, the crystallite has a compact structure and nucleation is a localized and rare event, i.e.~the induction time for crystal nucleation is long in comparison to the time needed by a particle to diffuse over its own diameter. This allows us to define a nucleation rate density, see FIG \ref{fig4}. For $P^*=40$, the crystal phase immediately forms a percolating network. Hence despite the approach to glassy dynamics, we observe the classical extremes of kinetics at a first order phase transition: nucleation and spinodal decomposition. (A similar observation has also been made for hard spheres \cite{Valeriani2012}.)

\begin{figure}[ht!]
\begin{center}
\includegraphics[height=5cm]{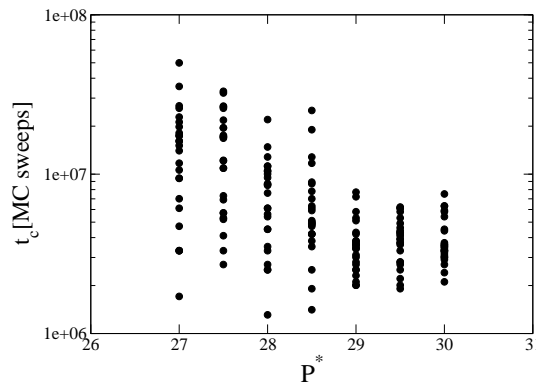}
\end{center}
\caption{\label{fig:scatter} Induction times versus pressure.}
\end{figure}

\begin{figure}[ht!]
\begin{center}
\includegraphics[height=5cm]{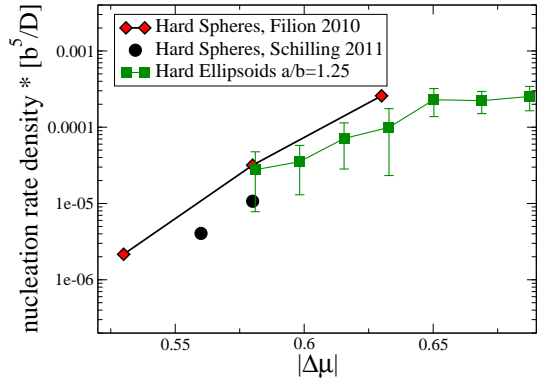}
\end{center}
\caption{\label{fig4} (a) Nucleation rate densities for hard spheres and ellipsoids as a function of the chemical potential difference $|\Delta\mu|$ between the meta-stable melt and the stable crystal phase. To allow for direct comparison with simulations with other types of local dynamics, the rate densities are multiplied by $b^5/D$, where D is the long-time self-diffusion constant. Data is averaged over 25 independent simulation runs and errorbars are calculated as the standard deviation.}
\end{figure}

For $P^*\leq30$, where the system crystallizes via nucleation, we have computed the nucleation rate density 
\[
I := \frac{1}{\langle V\rangle \langle t_c \rangle },
\]
where $\langle V\rangle$ is the mean volume of the system and $\langle t_c \rangle$ is the induction time averaged over 25 trajectories per value of pressure. 
As can be seen in FIG.\ref{time_evolution} (left) once a simulation run 
had produced a crystalline cluster consisting of 80 particles, it 
did not decay into the liquid state any more. Thus we used this value to 
identify 
an upper bound on the induction time. 

In FIG.~\ref{fig4} we compare the nucleation rate densities for ellipsoids to 
those for mono-disperse hard spheres (the hard sphere data was taken from refs.~\cite{Filion2010, Schilling2011}). 
At equal overcompression, the nucleation rate densities for the two systems coincide within the errorbars. We conclude that, at these moderate aspect ratios, the orientational degree of freedom of the ellipsoids does neither have a strong influence on the interfacial tension nor on the crystallization process. 
 
\subsection{Dynamic Heterogeneities and Crystallization}
\label{DynCryst}
In the rest of this paper, we focus on the dynamical structure of those regions in the melt that are about to crystallize.
From each trajectory we picked the first cluster of ellipsoids with $n_b>5$ that reached a size of 500 particles and studied the properties of these 500 particles backwards in time. The question we would like to address here, is whether we can see any unusual signature in the dynamics of these particles just before crystallization sets in when comparing them to the rest of the system. (We relaxed the criterion for crystallinity here from $n_b>8$ to $n_b>5$ in order to take particles on the surface of a cluster into account. This allows us to analyse our data also with respect to the surface effect discussed in ref.~\cite{Konishi2007, Sanz2011})

\begin{figure}[ht!]
\begin{center}
\includegraphics[height=5cm]{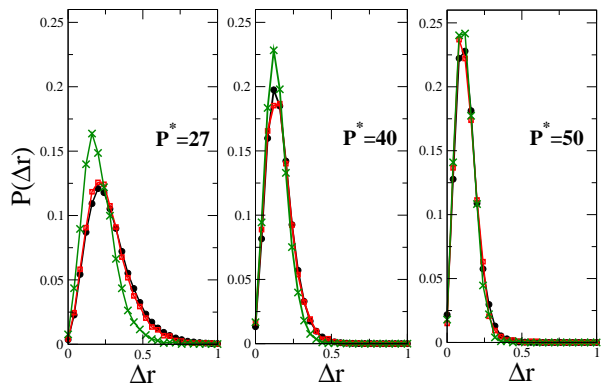}
\end{center}
\caption{\label{fig7} Mobility in the surrounding liquid (circles, black online) in comparison to the mobility of particles that are going to crystallize (squares, red online) and of the particles at the surface of the crystal, once the crystallite has formed (crosses, green online). The unit of length is $b$. 
}
\end{figure}

We define the mobility of a particle $i$ at a time $t$ as the distance 
\begin{equation}
\Delta r_i (t) = |\vec{r}_i(t) - \vec{r}_i(t-\Delta t)|
\label{def:mobil}
\end{equation} where
$\Delta t = 5\cdot 10^4$ MC sweeps. During this time interval, a particle 
travels on average a distance of $0.1 b $ to $0.25 b$ (depending on 
the pressure, see FIG.~\ref{fig1}). The average surface to surface distance 
between particles is slightly smaller than $0.1 b$. Hence the mobility 
defined in eqn.~\ref{def:mobil} captures information on the lengthscale that is relevant 
for local 
rearrangements of the fluid into the crystal.
FIG.~\ref{fig7} shows the probability distribution of $\Delta r_i (t)$.
Data for the surrounding liquid, which is not going to crystallize soon, is shown as circles (black online). The squares (red online) present the distribution of mobilities of those $500$ particles that are going to crystallize, taken just before the crystallization event. We identify the moment crystallization sets in by following the trajectories backwards in time from the time when there is a cluster of $500$ crystalline particles to the time when only 5\% of these particles are crystalline.  We use 5\% as a threshold, because below this number the probability of finding a crystalline particle among these $500$ is the same as the probability of finding a crystalline particle in the liquid.

There is no difference between the two distributions. The particles that will crystallize soon cannot be distinguished from the rest of the system. The hypothesis that the most mobile particles crystallize first, because they sample their phase space most rapidly, is not supported by our data.   

For hard spheres it has been reported that the mobility is enhanced at the surface of the crystalline cluster \cite{Konishi2007, Sanz2011}. To test for this effect, we computed the mobility of the surface particles once the crystallite had formed, i.e.~the mobility of those $100$ of the $500$ selected particles that had $n_b\le8$ when the other $400$ already had reached $n_b>8$ (FIG.~\ref{fig7}, crosses, green online). In contrast to refs.~\cite{Konishi2007, Sanz2011}, for low pressures we find a shift to lower mobilities, and at higher pressures no shift at all. 

\begin{figure}[ht]
\centering
  \includegraphics[height=5cm]{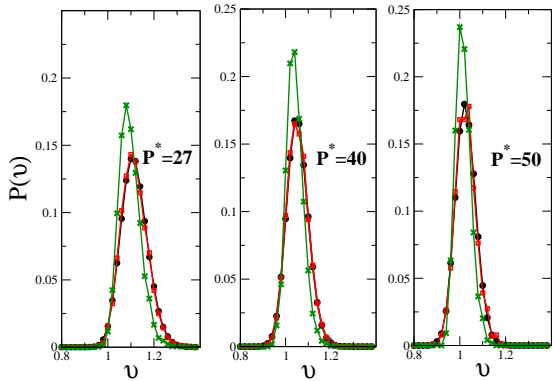}
  \caption{Single particle free volume in the surrounding liquid (circles, black online) in comparison to the single particle free volume of particles that are going to crystallize (squares, red online) and of the particles at the surface of the crystal, after the crystallite has formed (crosses, green online).  }
  \label{fig11}
\end{figure}

This observation is consistent with an analysis of the single particle free volume via Voronoi decomposition. FIG.~\ref{fig11} shows the distribution of the volumina of the Voronoi cells of all particles in the surrounding liquid (circles, black online) in comparison to those that are about to crystallize (squares, red online) and those that are at the surface of the crystal, once it has formed (crosses, green online). We observe no increase in the single particle free volume at the interface of the crystallite. There is no evidence of the modified crystallization process described in ref.~\cite{Sanz2011}.

\begin{figure}[ht!]
\begin{center}
\includegraphics[height=5cm]{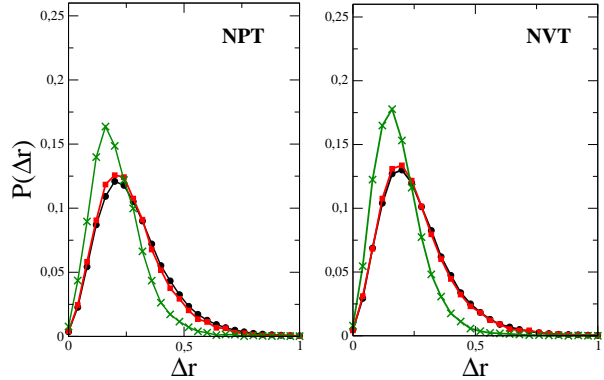}
\end{center}
\caption{\label{fig:compnvtnptmob} Comparison between simulation results for the NPT and the NVT ensemble: Mobilites at $P^*=27$ (left panel) and $\eta = 0.59$ (right panel). 
}
\end{figure}

\begin{figure}[ht!]
\begin{center}
\includegraphics[height=5cm]{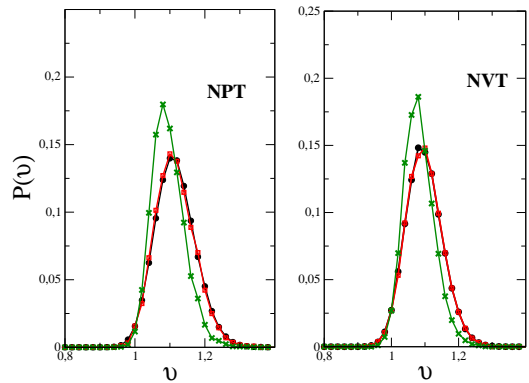}
\end{center}
\caption{\label{fig:compnvtnptvol} Comparison between simulation results for the NPT and the NVT ensemble: Single particle free volume at $P^*=27$ (left panel) and $\eta = 0.59$ (right panel). 
}
\end{figure}

One major difference between the simulations in ref.~\cite{Sanz2011} and our work is the choice of ensemble. To test whether the mobility at the interface is enhanced in the NVT ensemble, we repeated our analysis for $29$ trajectories at fixed $\eta =0.59$. In fig.~\ref{fig:compnvtnptmob} and fig.~\ref{fig:compnvtnptvol} we show the mobility distributions and the free volume distributions for NPT simulations at 
$P^*=27$ and for NVT simulations at $\eta =0.59$. There is no significant difference in any of the distributions. In particular, we do not see any enhanced free volume or mobility of the surface particles. In the NVT ensemble the liquid acts as a pressure reservoir on the particles that are going to crystallize. Hence the difference in observations regarding the free volume of particles at the cluster surface cannot be due to the choice of ensemble. It must be due to the different choice of model system (hard spheres in ref.~\cite{Sanz2011} and hard ellipsoids here). 

\begin{figure}[ht]
\centering
  \includegraphics[height=5cm]{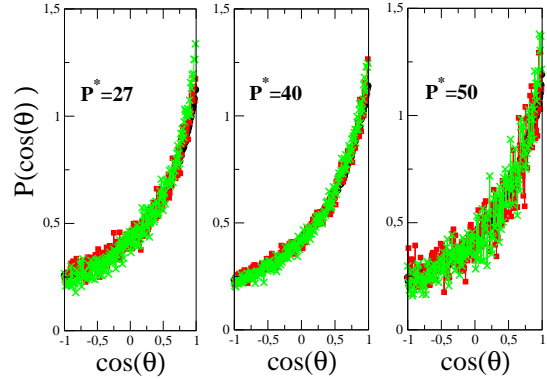}
  \caption{Alignment of motion of neighbouring particles in the surrounding liquid (circles, black online) in comparison to particles that are going to crystallize (squares, red online) and particles at the surface of the crystal, after the crystallite has formed (crosses, green online). See main text for definition of $\theta$.}
  \label{fig9}
\end{figure}

As the absolute distance traveled by a particle is not related to its likelyhood to crystallize, we now ask whether regions of orientationally correlated motion tend to crystallize faster than other regions. We define
\[
\cos(\theta) := \frac{\Delta \vec{r}_i(t)\cdot\Delta \vec{r}_j(t)}{|\Delta \vec{r}_i(t)|\cdot|\Delta \vec{r}_j(t)|}, 
\]
for neighbouring particles $i$ and $j$, where
\[
\Delta\ \vec{r}_i (t) = \vec{r}_i(t) - \vec{r}_i(t-\Delta t).
\]
$\cos(\theta)$ is presented in FIG.~\ref{fig9} with the same definitions of the symbols as before. Ellipsoids are considered to be neighbours if $\Delta r_{ij}<1.5b$. There is no difference between the three types of particles for any of the pressures. We conclude that neither regions of parallel motion nor regions where particles flow towards each other, are more likely to crystallize than other regions.

\begin{figure}[ht]
\centering
\includegraphics[height=5cm]{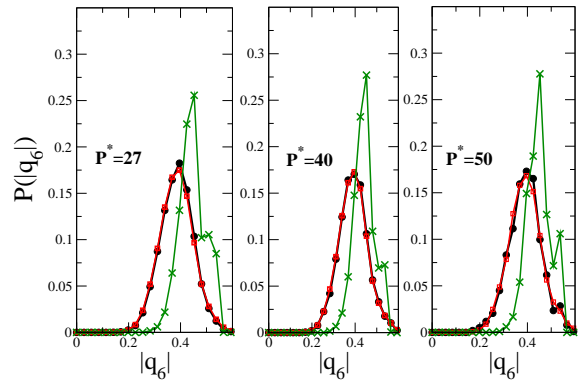}
\caption{ Distribution of $|q_6|$ values of particles in the surrounding liquid (circles, black online), of particles that are going to crystallize (squares, red online) and of the particles at the surface of the crystal, after the crystallite has formed (crosses, green online). 
}
\label{fig12}
\end{figure}

Finally we analyze the structural properties of the emerging crystallites in terms of the absolute value of the bond-orientation order-parameter $|q_6(i)|$ for each particle $i$, see Eq. \ref{BOP}.  FIG.~\ref{fig12} shows the distribution of $|q_6|$ for particles in the surrounding liquid (circles, black online), particles that are going to crystallize (squares, red online) and of the particles at the surface of the crystal, after the crystallite has formed (crosses, green online). Clearly the crystalline particles have a higher value of $|q_6(i)|$, but there is no difference between the particles that are going to crystallize and the surrounding liquid.

\section{Conclusion}
To summarize, we have studied crystallization in suspensions of hard ellipsoids at a moderate aspect ratio, $a/b=1.25$. We chose this system, because it is simple and monodisperse, its phase diagram is well known, it shows a glass transition and it can be realized experimentally\cite{Cohen2011,Mohraz2005, Zhang2010,Martchenko2011, Man2005,Kim2007,Crassous2012}. We pressure-quenched the system from the liquid state beyond the freezing pressure and studied its dynamical and structural properties during the subsequent crystallization process.
For moderate amplitudes of overcompression $P^*\leq30$, the system remained in a meta-stable melt state for long times and then crystallized via nucleation and growth. We showed that the nucleation rate densities for hard ellipsoids are consistent with those of monodisperse hard spheres (within the error-bars).
For the sub-diffusive, glass-like regime at high overcompressions, crystallization sets in on a time scale comparable to the relaxation times of the dynamic structure factor and particles diffuse less than their diameter. The crystalline regions form a percolating clusters. Hence, despite the approach to glassy dynamics, we observe the classical extreme cases of crystallization: nucleation and spinodal decomposition.

In order to test for correlations between dynamic heterogeneities and 
crystallization events 
we identified the regions that crystallized first and followed their behaviour 
backwards in time. We did not find any signature in the dynamic structure of 
the melt that would allow to predict which region was about to crystallize.

We neither saw enhanced moblities nor freeing up of volume at the 
melt/crystal interface. For high overcompression, ellipsoids close to the 
melt/crystal interface are as mobile as ellipsoids in the rest of the melt 
and for low overcompression they even slow down.  In addition, we tested for 
cooperative motion and did not find any signal that would allow to predict 
a crystallization site.

For our study, we conclude that we have not found any signature in the spatio-temporal structure of the supersaturated melt that would allow to predict imminent crystallization events.

\acknowledgements
We thank Chantal Valeriani for stimulating discussions.
This project has been financially supported by the DFG (SFB Tr6 and SPP1296) and by the National Research Fund, Luxembourg co-funded under the Marie Curie Actions of the European Commission (FP7-COFUND) and the National Research Fund Luxembourg under the project FRPTECD. Computer simulations presented in this paper were carried out using the HPC facility of the University of Luxembourg.


\end{document}